%% file: TOP-17-018_temp.tex
\pdfoutput=1

\documentclass[11pt,twoside,a4paper,cmspaper,final,collab]{cms-tdr}
\def\svnVersion{479235P}\def\cmsCernNoTag{CERN-EP-2018-074}\def\cmsCernDate{\today}\def\cmsMessage{Published in the Journal of High Energy Physics as \href{http://dx.doi.org/10.1007/JHEP10(2018)117}{\doi{10.1007/JHEP10(2018)117.}}}

\begin{document}\cmsNoteHeader{TOP-17-018}

\hyphenation{had-ron-i-za-tion}
\hyphenation{cal-or-i-me-ter}
\hyphenation{de-vices}
\RCS$HeadURL: svn+ssh://svn.cern.ch/reps/tdr2/papers/TOP-17-018/trunk/TOP-17-018.tex $
\RCS$Id: TOP-17-018.tex 479200 2018-10-25 07:52:31Z palencia $

\newlength\cmsFigWidth
\setlength\cmsFigWidth{0.4\textwidth}
\providecommand{\cmsLeft}{left\xspace}
\providecommand{\cmsRight}{right\xspace}
\providecommand{\CL}{CL\xspace}

\newlength\cmsTabSkip\setlength{\cmsTabSkip}{1ex}

\providecommand{\NA}{\ensuremath{\text{---}}}

\newcommand{\pT} {\pt}
\newcommand{\ppbar}{\ensuremath{\Pp\Pap}}
\newcommand{\roots}{\ensuremath{\sqrt{s}}}
\newcommand{\emu}{\ensuremath{\Pepm\PGmmp}\xspace}
\newcommand{\muObs}{\ensuremath{0.88 \pm 0.02\stat \pm 0.09\syst \pm 0.03\lum}\xspace}
\newcommand{\resultxsecmain}{\ensuremath{63.1 \pm 1.8\stat \pm 6.4\syst \pm 2.1\lum\unit{pb}}\xspace}
\newcommand{\tWxsectheo}{\ensuremath{\sigma_{\cPqt\PW}^\text{ref} = 71.7 \pm 1.8\,\text{(scale)} \pm 3.4\,\text{(PDF)}\unit{pb}}\xspace}

\newcommand{\lumivwunc}{\ensuremath{35.9\pm0.9\fbinv}}

\cmsNoteHeader{TOP-17-018}

\title{Measurement of the production cross section for single top quarks in association with \PW\ bosons in proton-proton collisions at $\sqrt{s}=13\TeV$}

\date{\today}

\abstract{
A measurement is presented of the associated production of a single top quark and a \PW\ boson in proton-proton collisions at $\roots= 13\TeV$ by the CMS Collaboration at the CERN LHC. The data collected corresponds to an integrated luminosity of 35.9\fbinv. The measurement is performed using events with one electron and one muon in the final state along with at least one jet originated from a bottom quark. A  multivariate discriminant, exploiting the kinematic properties of the events, is used to separate the signal from the dominant \ttbar background. The measured cross section of \resultxsecmain is in agreement with the standard model expectation.
}

\hypersetup{%
pdfauthor={CMS Collaboration},%
pdftitle={Measurement of the production cross section for single top quarks in association with W bosons in pp collisions at sqrt(s)=13 TeV},%
pdfsubject={CMS},
pdfkeywords={CMS, physics, top quarks}}

\maketitle

\section{Introduction }
\label{sec:introduction}
Single top quarks, observed for the first time by the D0~\cite{Abazov:2009ii} and CDF~\cite{Aaltonen:2009jj} Collaborations
at the Fermilab Tevatron, are produced via the electroweak interaction. There are three main production modes in proton-proton
(\Pp\Pp) or proton-antiproton (\ppbar) collisions: the exchange of a virtual \PW\ boson ($t$ channel), the production and decay
of a virtual \PW\ boson ($s$ channel), and the associated production of a top quark and a \PW\ boson (\cPqt\PW\ channel).

The \cPqt\PW\ process at the CERN LHC provides a unique opportunity to study the standard model (SM) and its extensions through the interference
of the process at next-to-leading order (NLO) with top quark pair (\ttbar) production~\cite{Frixione:2008yi,Belyaev:1998dn,White:2009yt}.
The \cPqt\PW\ process also plays an important role because of its sensitivity to the physics beyond the SM~\cite{Tait:2000sh,Cao:2007ea,Barger:2009ky}.

The \cPqt\PW\ production rate in \ppbar\ collisions at the Tevatron was negligible but at the LHC this process makes a significant
contribution to single top quark production. The CMS and ATLAS Collaborations have presented evidence for~\cite{Chatrchyan:2012zca, Aad:2012xca}
and observations of~\cite{Chatrchyan:2014tua,Aad:2015eto} this process in \Pp\Pp\ collisions at $\roots= 7$ and 8\TeV, respectively. The ATLAS
Collaboration has also measured the production cross section using 13\TeV data~\cite{Aaboud:2016lpj}.

\begin{figure}[htbp!]
  \centering
    \includegraphics[width=0.35\textwidth]{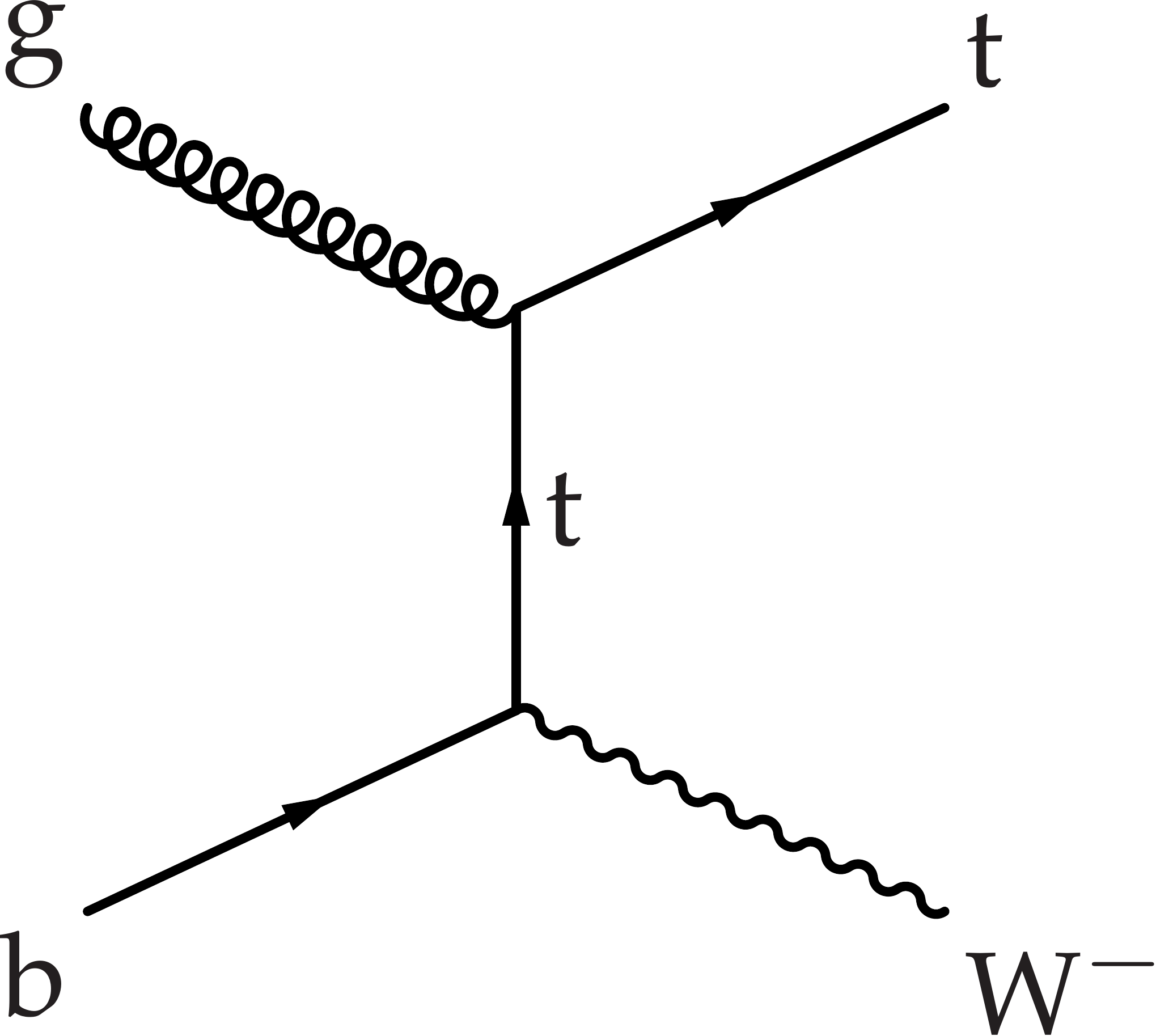}
    \hspace{1cm}
    \includegraphics[width=0.35\textwidth]{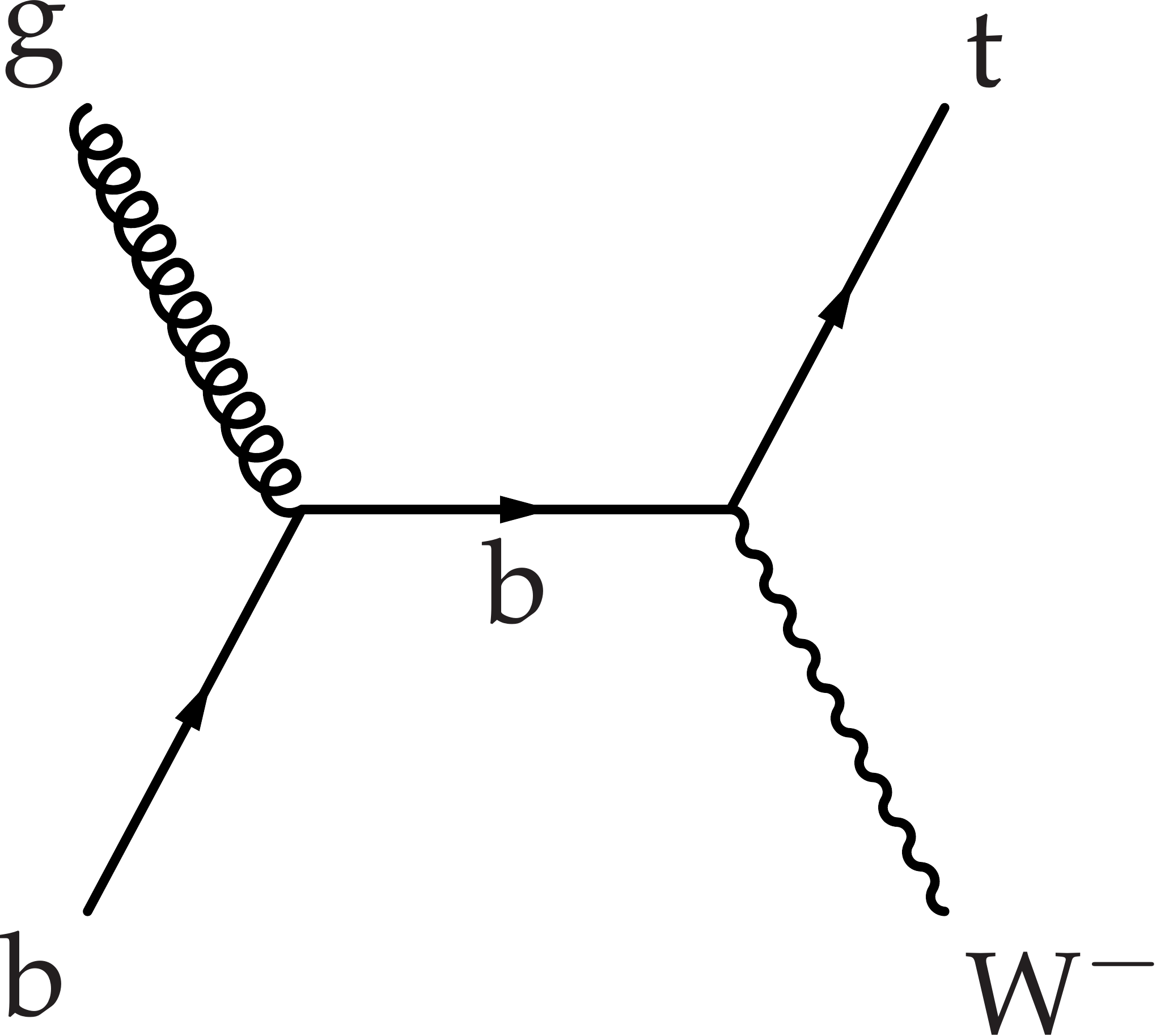}
    \caption{Leading-order Feynman diagrams for single top quark production in the \cPqt\PW\ channel that implicitly include
the charge-conjugate contributions.}
      \label{fig:fig_1}
\end{figure}

The \cPqt\PW\ production cross section is computed at an approximate next-to-next-to-leading order (NNLO). The corresponding theoretical
prediction for the \cPqt\PW\ cross section in \Pp\Pp\ collisions at $\roots= 13\TeV$, assuming a top quark mass ($m_\cPqt$) of 172.5\GeV, is
\tWxsectheo~\cite{Kidonakis:2015nna}. The first uncertainty refers to the factorization ($\mu_\mathrm{F}$) and renormalization
($\mu_\mathrm{R}$) scales in quantum chromodynamics (QCD), and the second to parton distribution functions (PDFs). The quoted cross section
includes the charge-conjugate modes. The leading-order (LO) Feynman diagrams for \cPqt\PW\ production are shown in Fig.~\ref{fig:fig_1}.

This paper reports the first measurement from the CMS Collaboration of \cPqt\PW\ production in \Pp\Pp\ collisions at $\roots= 13\TeV$. The measurement
uses data recorded by CMS during 2016, corresponding to an integrated luminosity of \lumivwunc.
The analysis is performed using the \emu dilepton channel, in which both \PW\ bosons, either produced in association with
the top quark or from the decay of the top quark, decay
leptonically into a muon or an electron ($\ell$), and a neutrino.
Events with \PW\ bosons decaying into \Pgt\ leptons that decay into electrons or muons also contribute to the measurement.
The primary background to \cPqt\PW\ production in this final state comes from \ttbar
production, with Drell--Yan (DY) production of \Pgt\ lepton pairs that decay leptonically being the next most significant background.
To extract the signal, the analysis uses a multivariate technique,
exploiting kinematic observables to distinguish the \cPqt\PW\ signal from the dominant \ttbar background.

The paper is structured as follows. Section~\ref{sec:samples} provides a summary of the CMS detector and of the
Monte Carlo (MC) event simulation. The object and event selection criteria are discussed in Section~\ref{sec:eventselection}.
The description of the method used to separate the \cPqt\PW\ signal from the \ttbar background is given in Section~\ref{sec:res}. The
sources of systematic uncertainties are discussed in Section~\ref{sec:syst}. The extraction of the \cPqt\PW\ production cross section
is described in Section~\ref{sec:fit}, and a summary of the results is presented in Section~\ref{sec:summary}.

\section{The CMS detector and Monte Carlo simulation}
\label{sec:samples}
The CMS detector has a superconducting solenoid in its central region that provides an axial magnetic field
of 3.8\unit{T}. The silicon pixel and strip trackers cover $0 < \phi <2\pi$ in azimuth and $\abs{\eta}<2.5$
in pseudorapidity. The lead tungstate crystal electromagnetic calorimeter, and the brass and scintillator
hadron calorimeter are located inside the solenoid. These are used to identify electrons,
photons, and jets. Muons are measured in gas-ionization detectors embedded in the steel
flux-return yoke outside the solenoid. The detector is nearly hermetic, providing reliable
measurement of the momentum imbalance in the plane transverse to the beams. A two-level
trigger system selects the most interesting \Pp\Pp\ collisions for offline analysis.
A more detailed description of the CMS detector, together with a definition
of the coordinate system used and the relevant kinematic variables, can be
found in Ref.~\cite{Chatrchyan:2008zzk}.

The \cPqt\PW\ signal is simulated at NLO using \POWHEG v1~\cite{bib:powheg3} with
the NNPDF 3.0 PDF set~\cite{Ball:2014uwa}, and \PYTHIA~v8.205~\cite{Sjostrand:2006za,Sjostrand:2014zea},
with the underlying event tune CUETP8M1~\cite{Khachatryan:2015pea,Skands:2014pea}, is used for parton showering and hadronization.
At NLO in perturbative QCD, \cPqt\PW\ production interferes with \ttbar production~\cite{Frixione:2008yi,Belyaev:1998dn,White:2009yt}.
Two different procedures can be used to account for this interference:
the "diagram removal" (DR)~\cite{Frixione:2008yi} approach, where all NLO diagrams that are doubly resonant
are excluded from the signal definition; and the "diagram subtraction" (DS)~\cite{Frixione:2008yi,PhysRevD.61.034001} approach, in which
the differential cross section is modified with a gauge-invariant subtraction term, which locally cancels the contribution of
\ttbar diagrams. The DR scheme is
used here, and it has been verified that the number of predicted events after the full selection is comparable
with that obtained from the DS approach.

The NLO \POWHEG v2~\cite{bib:powheg2} setup is used to simulate \ttbar events, as well as
the dependency of the \ttbar production on $m_\cPqt$,
$\mu_\mathrm{R}$ and $\mu_\mathrm{F}$,
and the PDF set. The NNPDF 3.0 set is used as the default PDF set.
Parton showering and hadronization for the \ttbar events are handled by
\PYTHIA~v8.205  with the underlying event tune CUETP8M2T4~\cite{CMS-PAS-TOP-16-021}.

Background contributions from processes other than \ttbar are also estimated from MC simulations.
The DY and \PW+jets background samples are generated at NLO  with \MGvATNLO v2.2.2~\cite{Alwall:2014hca} with NNPDF 3.0 PDFs,
interfaced with \PYTHIA~v8.205, with the CUETP8M1 underlying event tune for fragmentation and hadronization.
These processes are simulated with up to two additional partons and the FxFx scheme~\cite{Frederix:2012ps} is used for the merging.
The contributions from \PW\PW, \PW\cPZ, and \cPZ\cPZ\ (referred to as VV) processes
are simulated at LO with \PYTHIA~v8.205 with the CUETP8M1 underlying event tune.
Other contributions from \PW\ and \cPZ\ boson production in association with \ttbar events
(referred to as $\ttbar$V) are simulated at NLO using \MGvATNLO v2.2.2 and interfaced with \PYTHIA~v8.205 with the CUETP8M1 underlying event tune.
Finally, the \ttbar and \PW+jets samples described above, in the lepton+jets final state, are used to
determine the contribution to the background from events with a jet incorrectly reconstructed as a lepton or
with a lepton incorrectly identified as being isolated.
These last contributions to the background are
labeled non-\PW/\cPZ\ as they contain a lepton candidate that does not originate from a leptonic decay of a gauge boson.

For comparison with the measured distributions, the event yields in the simulated samples are normalized using
the integrated luminosity and their theoretical cross sections. These are taken from NNLO (\PW+jets and
DY~\cite{PhysRevD.86.094034}), approximate NNLO (single top quark \cPqt\PW\ channel~\cite{Kidonakis:2015nna}),
and NLO (diboson~\cite{bib:mcfm:diboson}) calculations.
For the simulated \ttbar sample, the full NNLO plus next-to-next-to-leading-logarithmic accuracy calculation~\cite{Czakon:2013goa},
performed with the \textsc{Top++} 2.0 program~\cite{Czakon:2011xx}, is used.
The PDF uncertainty is added in quadrature to the uncertainty associated with the strong coupling constant (\alpS)
to obtain a \ttbar production cross section of $832\,^{+20}_{-29}\,\text{(scale)}\pm 35\,\text{(PDF+\alpS)}\unit{pb}$ assuming
$m_\cPqt = 172.5\GeV$. The simulated samples include additional interactions per bunch crossing (pileup), with the distribution
matching that observed in data, with an average of 23 collisions per bunch crossing.

\section{Event selection}
\label{sec:eventselection}
In the SM, a top quark decays almost exclusively into a \PW\ boson and a bottom quark. The analysis
uses the \emu decay channel, in which the \PW\ boson produced in association with
the top quark and the \PW\ boson from the decay of the top quark both decay leptonically,
one into an electron and the corresponding neutrino, and the other into a muon and the corresponding neutrino.
This leads to a final state composed of two oppositely charged
leptons, a jet resulting from the fragmentation of a bottom quark, and two neutrinos.
The event selection described here follows closely that used in the measurement of the top quark-antiquark
pair production cross section in the dilepton channel~\cite{CMS-PAS-TOP-16-005}.

Events are required to pass either a dilepton or single-lepton trigger. The dilepton triggers require events
to contain either one electron with transverse momentum $\pt > 12\GeV$
and one muon with $\pt > 23\GeV$, or one electron having $\pt > 23\GeV$ and one muon with $\pt > 8\GeV$.
In addition, single-lepton triggers with one electron (muon) with $\pt > 27\,(24)\GeV$ are used to increase the efficiency.
The efficiency for the combination of the single-lepton and dilepton triggers is measured in data events passing the dilepton
selection criteria given below and collected using triggers based on the \pt imbalance in the event. This efficiency is found to be $\approx$98\%.
The efficiency of the simulated trigger is corrected to match that observed in data using a multiplicative scale factor (SF).

The particle-flow (PF) algorithm~\cite{Sirunyan:2017ulk} attempts to reconstruct and identify each individual particle in an event
with an optimized combination of information from the various elements of the CMS detector. Leptons (electrons~\cite{Khachatryan:2015hwa}
or muons~\cite{Chatrchyan:2012xi}) in the event are required to be well isolated and to have $\pt > 20\GeV$ and $\abs{\eta} < 2.4$.
Isolation requirements are based on the scalar sum of the \pt of all PF candidates, reconstructed inside a cone of
$\Delta R = \sqrt{\smash[b]{(\Delta\eta)^2+(\Delta\phi)^2}} = 0.3$ (0.4) centered on the electron (muon), excluding the contribution from
the lepton candidate. Tracks not coming from the main vertex are excluded in the calculation. This isolation variable is required to be
smaller than 6 (15)\% of the electron (muon) \pt. Events with \PW\ bosons decaying into \Pgt\ leptons contribute to the measurement
only if the \Pgt\ leptons decay into electrons or muons that satisfy the selection requirements. In events with more than two leptons passing
the selection, the two with the largest \pt are selected for further study.

Jets are reconstructed from the PF candidates using the anti-\kt clustering algorithm~\cite{Cacciari:2008gp, Cacciari:2011ma} with a distance
parameter of 0.4.
The jet momentum is determined as the vectorial sum of all particle momenta in the jet, and on average is found from simulation to be within 5 to 10\% of the true
momentum over the whole \pt spectrum and detector acceptance. Additional \Pp\Pp\ interactions within the same or nearby bunch crossings can
contribute additional tracks and calorimetric energy depositions to the jet momentum. To mitigate this effect, tracks identified as originating
from pileup vertices are discarded, and an offset correction is applied to correct for the remaining contributions.
Jet energy corrections, derived from simulation, are applied so that the average response to jets matches the particle level jets~\cite{CMS-PAS-JME-16-003}.
In situ measurements of the momentum balance in dijet,
photon+jet, \cPZ+jet, and multijet events are used to account for any residual differences in jet energy scale (JES) between data and simulation. Additional
selection criteria are applied to each jet to remove jets potentially dominated by anomalous contributions from various subdetector components or
reconstruction failures.
Jets are required to have $\pt > 30\GeV$ and $\abs{\eta} < 2.4$. In order to avoid double counting, jets within a cone of $\Delta R = 0.4$ with respect
to the selected leptons are not considered.
Jets passing the above identification criteria but with \pt between 20 and 30\GeV are referred to as "loose jets".

The missing transverse momentum vector \ptvecmiss is defined as the negative vector sum of the momenta of all reconstructed PF candidates in an event,
projected onto the plane perpendicular to the direction of the beam axis.
Its magnitude is referred to as \ptmiss and the corrections to jet momenta are propagated to the \ptmiss\ calculation~\cite{CMS-PAS-JME-16-004}.

In contrast to some sources of backgrounds, such as DY events, the \cPqt\PW\ final state contains a bottom quark. The identification of jets originating
from \cPqb\ quarks results in a significant reduction in background.
Jets are identified as \cPqb\ jets using the combined secondary vertex algorithm v2~\cite{Sirunyan:2017ezt},
with an operating point that yields identification efficiencies of $\approx$70\% and misidentification (mistag)
probabilities of about 1\% and 15\%~\cite{Sirunyan:2017ezt} for light-flavor jets (\cPqu, \cPqd, \cPqs,
and gluons) and \cPqc\ jets, respectively, as estimated from simulated events.

\begin{figure}[htbp!]
    \includegraphics[width=0.49\textwidth]{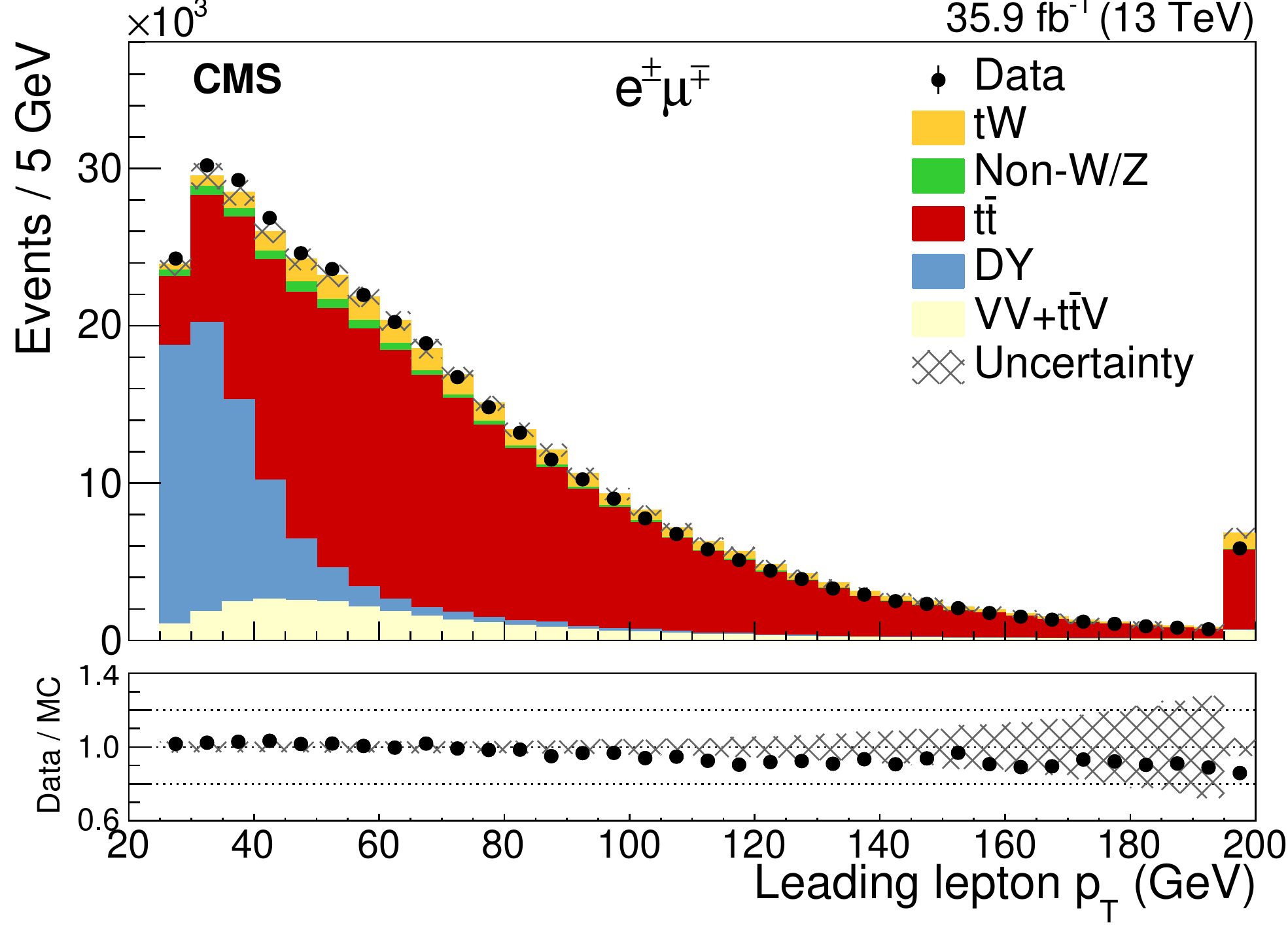}
    \includegraphics[width=0.49\textwidth]{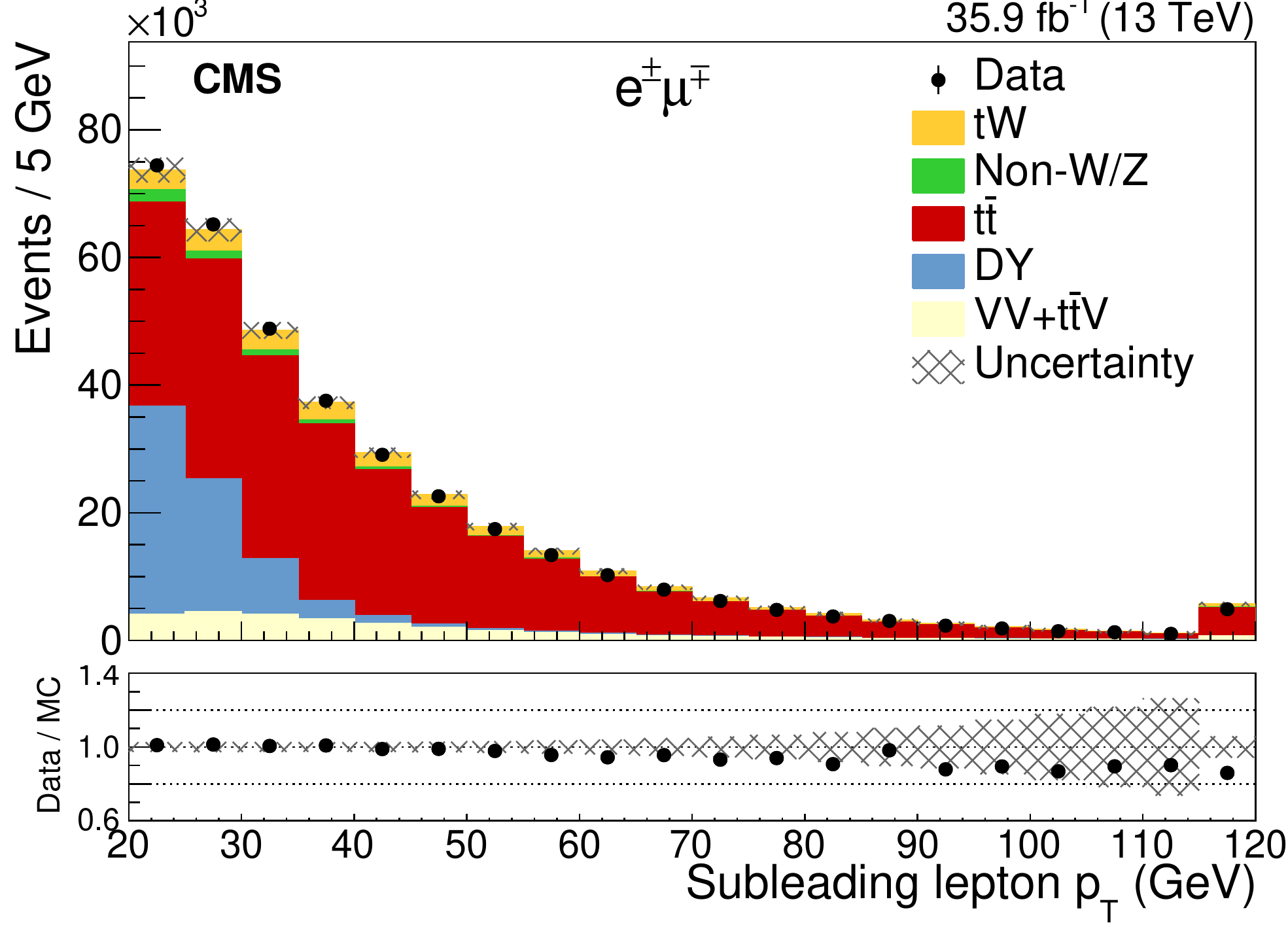} \\
    \includegraphics[width=0.49\textwidth]{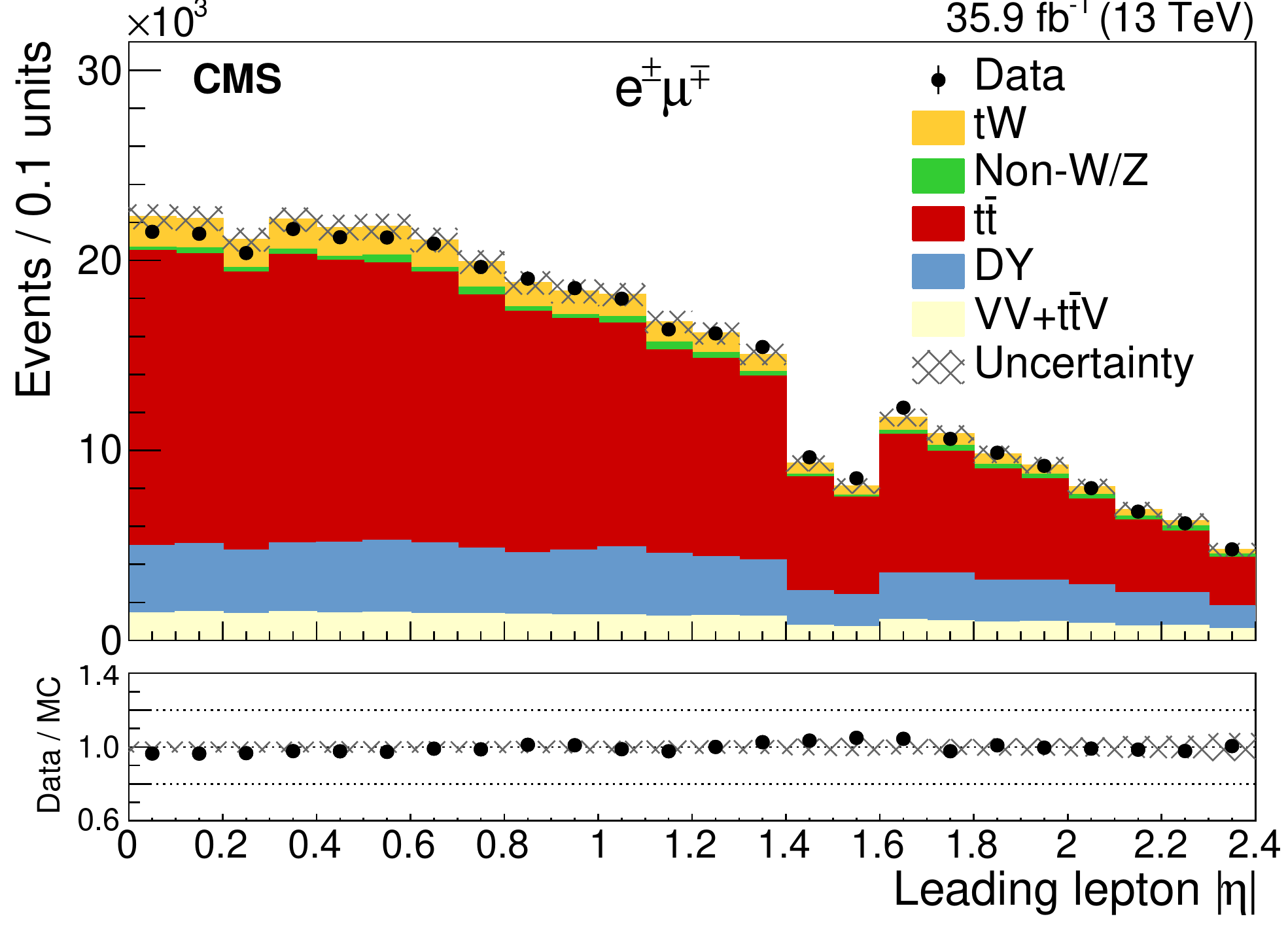}
    \includegraphics[width=0.49\textwidth]{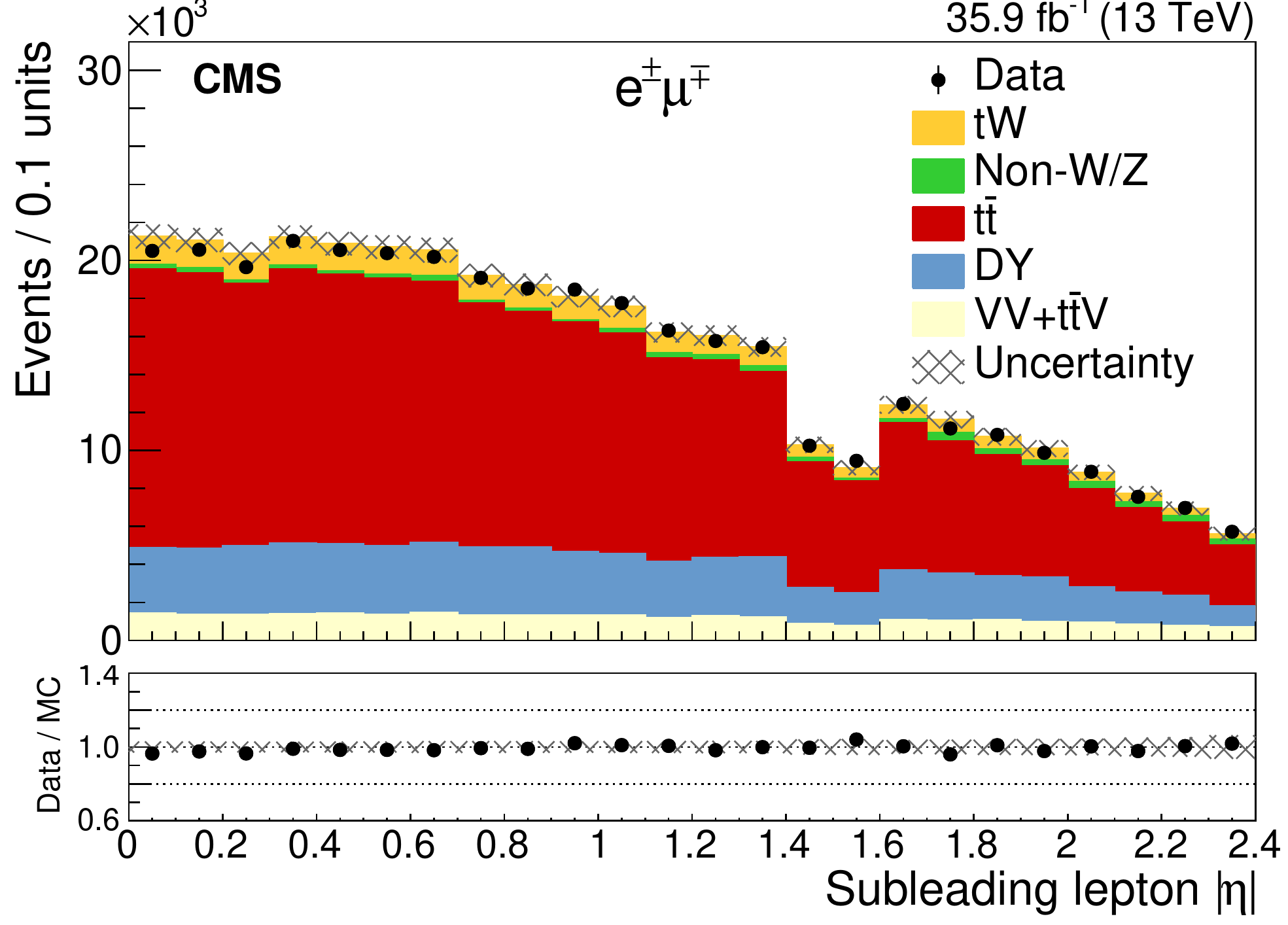} \\
    \includegraphics[width=0.49\textwidth]{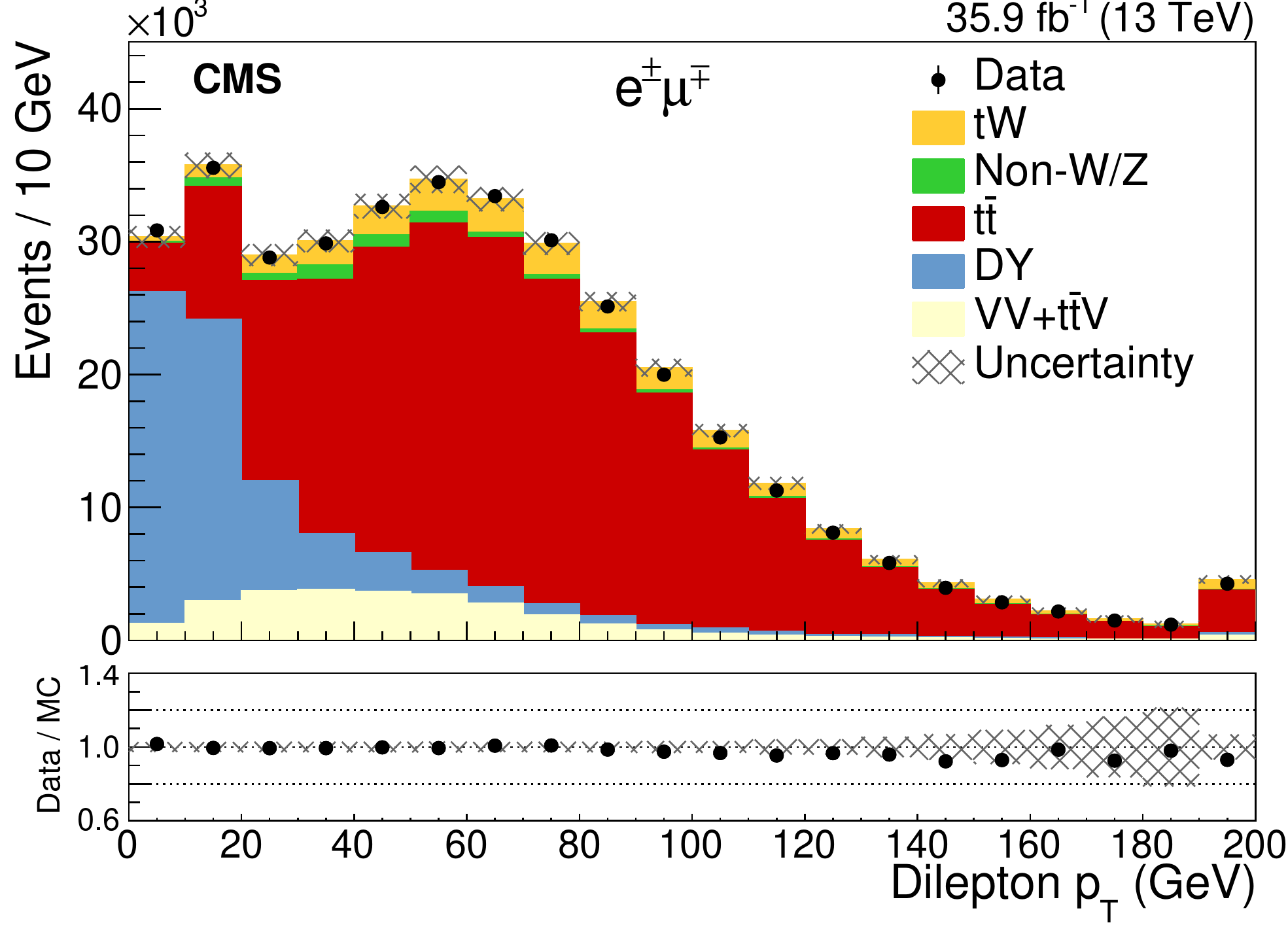}
    \includegraphics[width=0.49\textwidth]{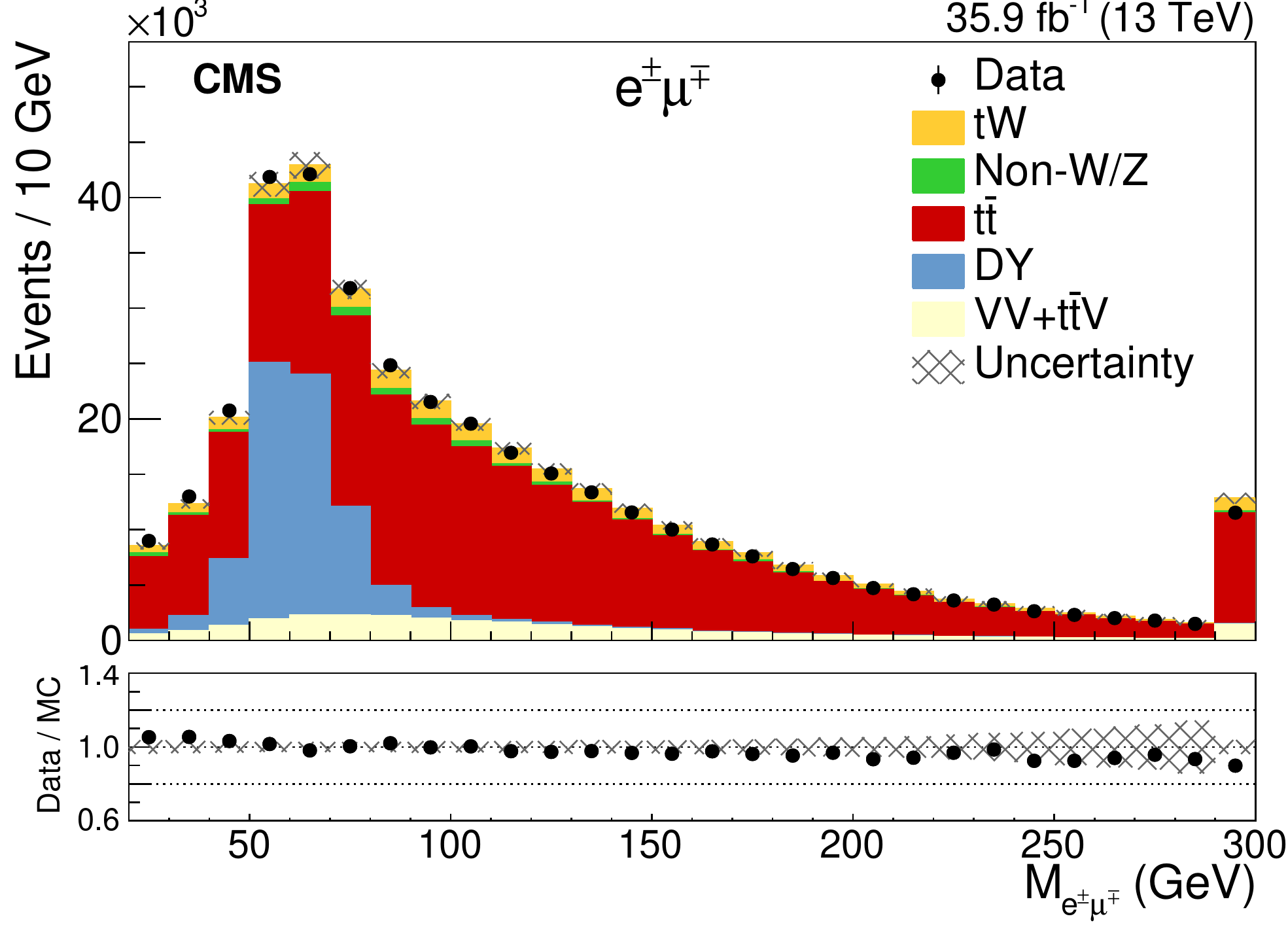} \\
   \caption{Comparison of several lepton kinematic variables for the observed data and simulated events after the dilepton
   selection is applied. The last bin includes overflow events. The error band includes the statistical and all systematic uncertainties, except those from background
   normalization. The bottom of each panel shows the ratios of data to the sum of the expected yields.}
    \label{fig:dataMCdilepton1}
\end{figure}

Events are classified as belonging to the \emu final state if the two leptons with larger \pt (leading leptons)
passing the above selection criteria are an electron and a muon of opposite charge. We require the leading lepton
to have $\pT > 25\GeV$. As this requirement for electrons is lower than the corresponding trigger threshold, some
of the phase space is triggered only by the muon or dilepton triggers. This effect is taken into account in the
measurement of the trigger efficiency. To reduce the contamination from DY production of \Pgt\ lepton pairs with
low invariant dilepton mass, the invariant mass of the lepton pair is required to be greater than 20\GeV.
Figure~\ref{fig:dataMCdilepton1} shows a comparison of several lepton kinematic distributions in data and simulated events
after this baseline selection.

\begin{figure}[htbp!]
  \centering
    \includegraphics[width=0.49\textwidth]{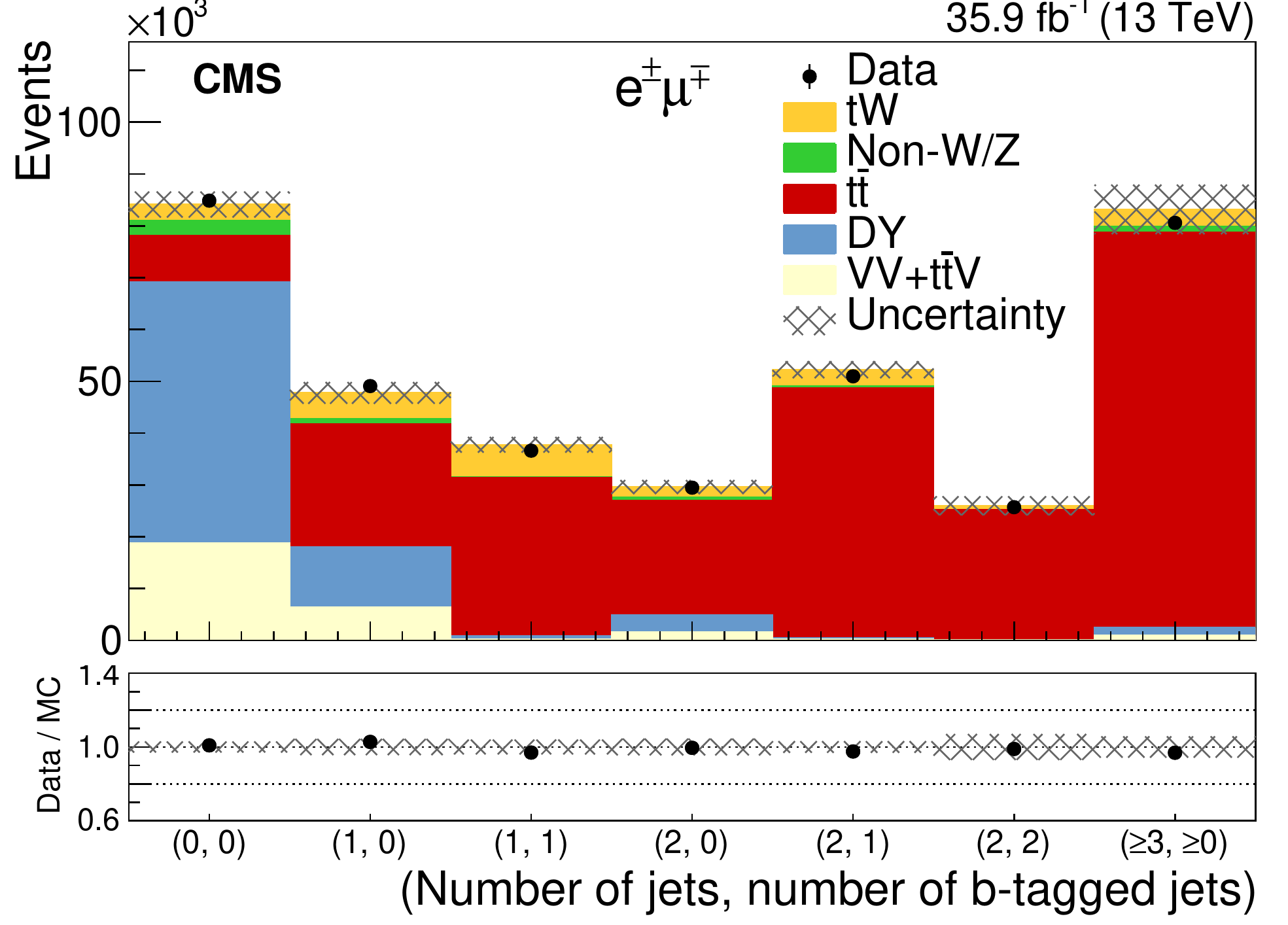}
    \caption{Yields observed in data, compared with those expected from simulation, as a function of the number of jets and number of \cPqb-tagged jets
     for events passing the baseline dilepton selection. The error band includes the statistical and all systematic uncertainties, except those
     from background normalization. The bottom of each panel shows the ratios of data to the sum of the expected yields.}
    \label{fig:njetsnbtags}
\end{figure}

Figure~\ref{fig:njetsnbtags} shows a comparison of the yields observed in data with those estimated from simulated events, classified
according to the number of jets and identified \cPqb\ jets in the event. As expected, the most signal-enriched region is the one with one jet
that is tagged as a bottom jet (1j1b region), but the size of the signal in comparison with the overwhelming \ttbar background
makes a cut-based analysis extremely challenging. Therefore, a multivariate analysis is pursued.

For the final analysis, the events are classified into three independent categories: a signal-enriched region with 1j1b events, and two
background-dominated regions with two jets, one with one \cPqb-tagged jet (2j1b) and one with two \cPqb-tagged jets (2j2b).

\section{Signal extraction}
\label{sec:res}
As noted previously, following the baseline event selection the data sample in the
1j1b region consists primarily of \ttbar events
with a significant number of \cPqt\PW\ signal events (as can be seen from Fig.~\ref{fig:njetsnbtags}). Given that there is no single observable that clearly discriminates
between the signal and background, a multivariate method is used to discriminate the \cPqt\PW\ signal from the main
background process, \ttbar. Several observables are combined into a single discriminator using a boosted decision tree (BDT) technique~\cite{Hocker:2007ht,Yang:2005nz}.
In this analysis, the BDT implementation is provided by the "Toolkit
for Multivariate Data Analysis"~\cite{Hocker:2007ht} package, using the gradient boost algorithm~\cite{Hocker:2007ht,Yang:2005nz}. The training of the BDT is performed
using dedicated simulated samples for \cPqt\PW\ and \ttbar\ that are statistically independent from those used for the signal extraction.
The input variables used for training the BDT in the 1j1b region, listed in order of importance to the BDT training are shown below. The order of importance is determined by counting how often each variable is used to split decision trees. The counts are weighted by the separation gain squared achieved by the variable and by the number of events in the node.

\begin{itemize}
\item \pt of leading loose jet, set to 0 for events with no loose jets present;
\item magnitude of the vector sum of the \pt's of leptons, jet, and \ptvecmiss\ ($\pt^\text{sys}$);
\item \pt of the jet;
\item ratio of the scalar sum of the \pt of the leptons to the scalar sum (\HT) of the \pt's of leptons, jet, and \ptmiss;
\item number of loose jets;
\item centrality (ratio between the scalar sums of the \pt and of the total momentum) of the jet and the two leptons;
\item magnitude of the vector sum of the \pt of the jet and leptons;
\item \HT;
\item ratio of $\pt^\text{sys}$ to \HT for the event;
\item invariant mass of the combination of the leptons, jet, and \ptmiss;
\item number of \cPqb-tagged loose jets.
\end{itemize}

\begin{figure}[htbp!]
  \centering
    \includegraphics[width=0.49\textwidth]{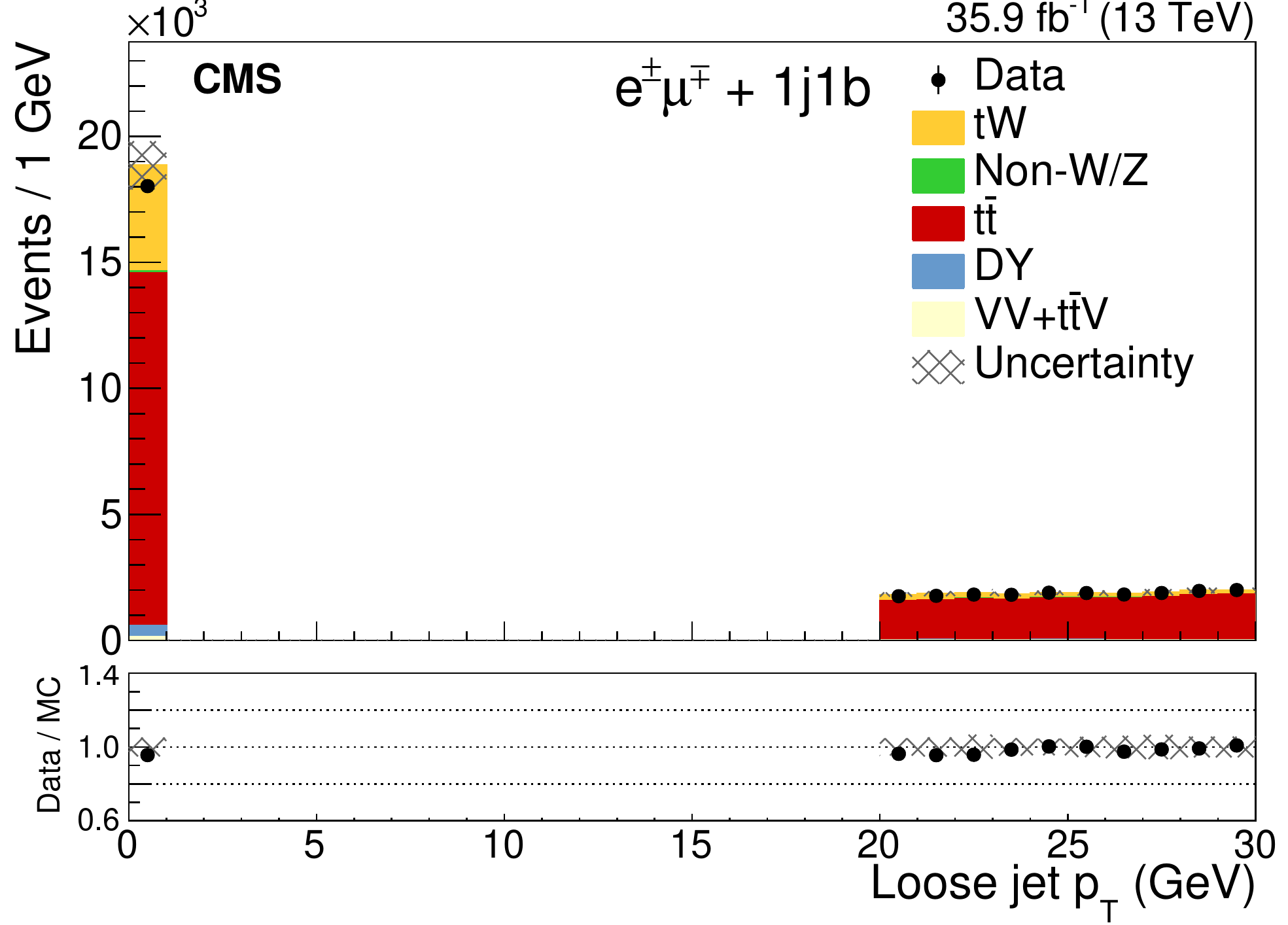}
    \includegraphics[width=0.49\textwidth]{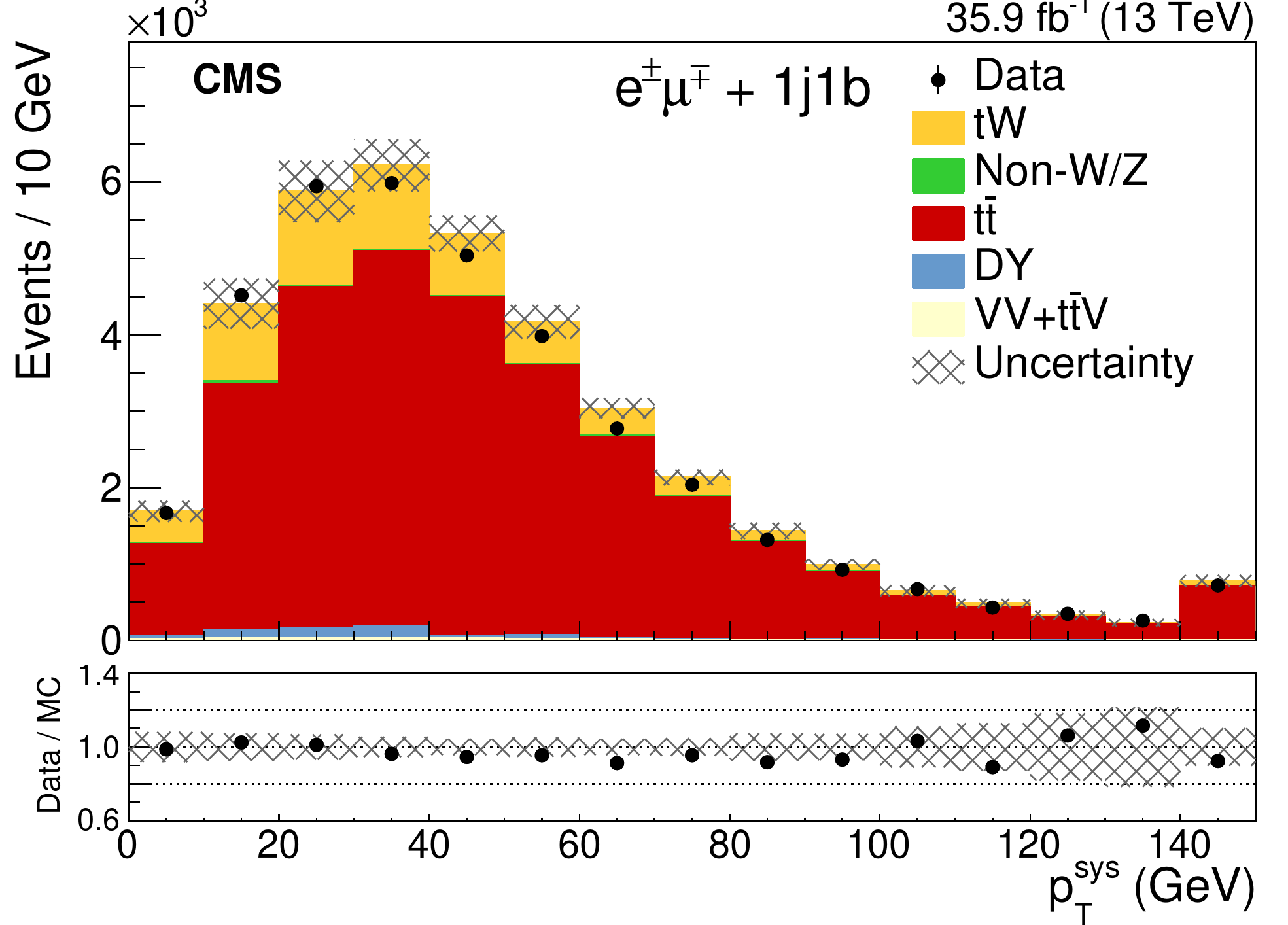} \\
    \includegraphics[width=0.49\textwidth]{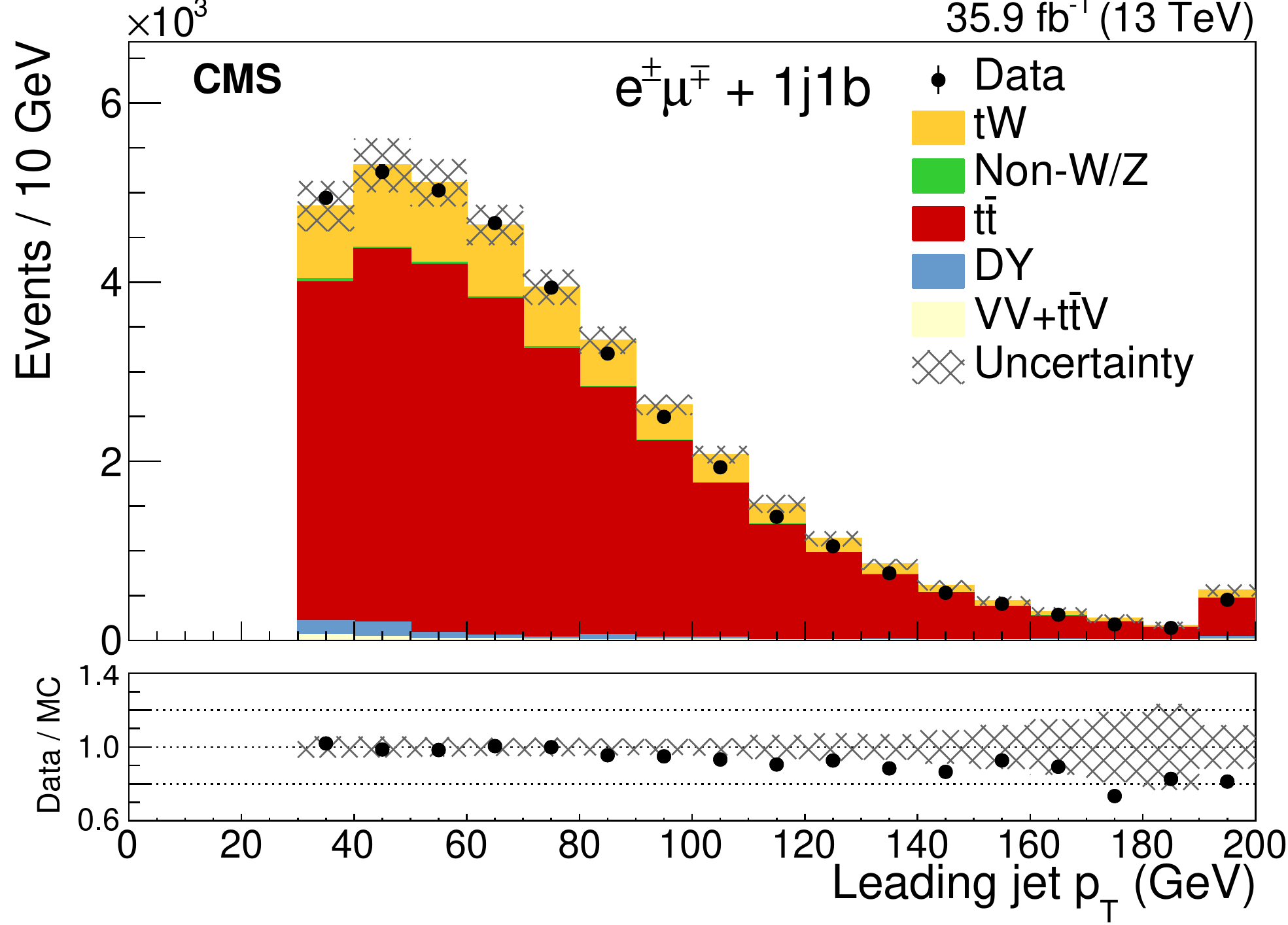}
    \includegraphics[width=0.49\textwidth]{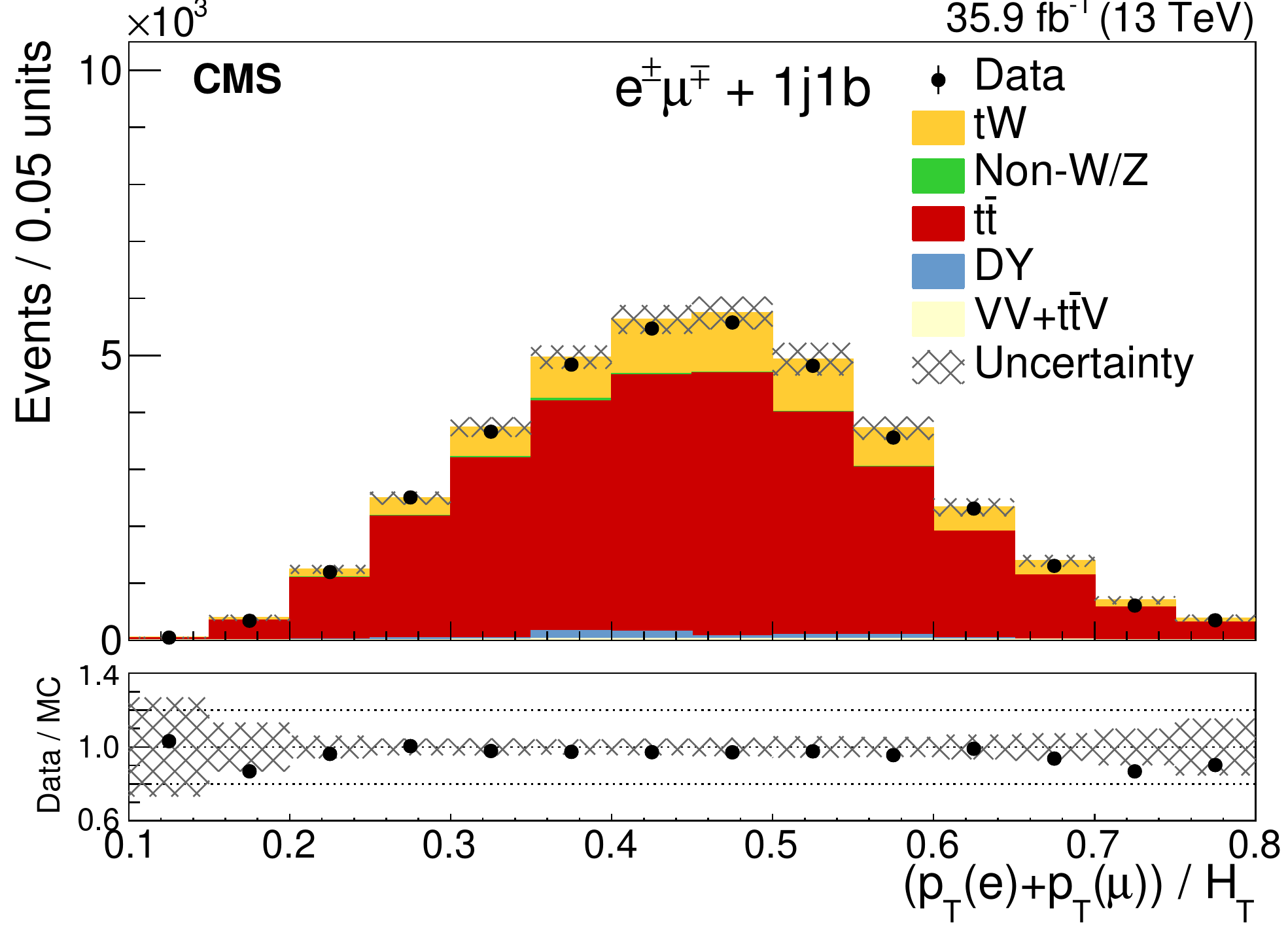} \\
      \caption{Most discriminating variables used for the training of the BDT in the 1j1b category. The last bin includes overflow events. The error band includes the statistical
      and all systematic uncertainties. The bottom of each panel shows the ratios
      of data to the sum of the expected yields.}
      \label{fig:fig_supMat1}
\end{figure}

The distributions of the four variables with the most discriminating power, in data and simulated events, are shown in Fig.~\ref{fig:fig_supMat1}.

A separate BDT is trained with events in the 2j1b region.
The input variables used for the training, listed in order of importance to the BDT training, are the following:

\begin{itemize}
\item separation in the $\phi-\eta$ space between the dilepton and dijet systems, $\Delta R$ (\emu, j$_1$j$_2$);
\item separation in the $\phi-\eta$ space between the dilepton system and the dijet and \ptmiss system, $\Delta R$ (\emu, j$_1$j$_2$\ptmiss);
\item \pt of the subleading jet;
\item separation in the $\phi-\eta$ space between the leading lepton and the leading jet, $\Delta R$ ($\ell_1$, j$_1$).
\end{itemize}

The 2j2b control region is highly enriched with \ttbar events and is used to constrain this main source of background using the
\pt distribution of the subleading jet.
This variable is sensitive to JES variations and, therefore, useful in constraining this source of systematic uncertainty.

The signal is extracted by performing a maximum likelihood fit to one measured distribution in each of the three measurement
regions: the distributions of the BDT output in the 1j1b and 2j1b categories, and of the \pt of the subleading jet
in the 2j2b region. The binning of the BDT outputs is chosen such that each bin contains approximately the same number
of \ttbar background events.
This selection of binning ensures that enough background events populate all the bins of the distribution,
helping to constrain the systematic uncertainties. The fit is performed simultaneously in the three regions. The 
uncertainties on the \ttbar overall normalization and shapes (including migrations into/out of the signal and control regions) are handled using
different nuisance parameters, one for each systematic uncertainty and for all regions.

The likelihood used in this statistical analysis, $\mathcal{L}(\mu,\vec{\theta})$,
is a function of the signal strength, defined as the ratio of measured and expected cross
sections $\mu = \sigma_{\cPqt\PW}/\sigma_{\cPqt\PW}^{\text{ref}}$, and a
set of nuisance parameters, $\vec{\theta}$, that parametrize the systematic uncertainties
present in the analysis.
The expected numbers of both signal and background events in each bin of the distributions are obtained
using normalized distributions (templates) from simulation, and are a function of $\vec{\theta}$ and,
in the case of the signal, $\mu$. The likelihood function is constructed as the product of Poisson
probabilities, corresponding to the number of events in each bin of the distributions. Additionally, the
systematic uncertainties are introduced in the likelihood by multiplying it by the prior of each nuisance
parameter, which are log-normal probability density functions.

The best value for $\mu$ is obtained by maximizing the likelihood function with respect to all its parameters.
The 68\% confidence interval is obtained by considering variations of the test statistic used in
Ref.~\cite{Khachatryan:2014jba} by one unit from its minimum.

\section{Systematic uncertainties}
\label{sec:syst}

The measurement of the \cPqt\PW\ production cross section is affected by systematic uncertainties that
originate from detector effects and event modeling, which can change the shape and/or the normalization
of the distributions used in the fit.
Each source of systematic uncertainty is assessed individually by appropriate variations of the MC
simulations or by variations of parameter values in the analysis within their estimated uncertainties, and propagated to the signal strength.
A nuisance parameter represents each of the sources and these parameters are used, together with the \cPqt\PW\ production
cross section, as parameters in the fit.

\subsection{Experimental uncertainties}
The following sources of experimental uncertainty are considered in the analysis:

\begin{itemize}
   \item The uncertainties in the trigger and lepton identification efficiencies in simulation are estimated
   by varying data-to-simulation SFs by their uncertainties. These are about 0.7 and 1.5\%, respectively,
   with some dependence on the lepton \pt and $\eta$. For lepton efficiencies we have two nuisance parameters,
   one for electrons and one for muons.
   \item The uncertainty due to the limited knowledge of the JES and jet energy resolution is determined by
varying the scale and resolution within the uncertainties in bins of \pt and $\eta$, typically by a few percent~\cite{CMS-PAS-JME-16-003}.
JES uncertainties are propagated to \ptvecmiss.
   \item The uncertainties resulting from the \cPqb\ tagging efficiency and misidentification rate
are determined by varying, within their uncertainties, the \cPqb\ tagging data-to-simulation SFs of the \cPqb\ jets and the light-flavor jets, respectively.
These uncertainties depend on the \pt and $\eta$ of the jet and amount to approximately 2\% for \cPqb\ jets and 10\% for
mistagged jets~\cite{Sirunyan:2017ezt}, as determined in simulated \ttbar events.
   \item The uncertainty assigned to the number of pileup events in simulation is obtained by changing the
inelastic \Pp\Pp\ cross section, which is used to estimate the pileup in data, by
$\pm$4.6\%~\cite{Aaboud:2016mmw}.
   \item The uncertainty in the integrated luminosity is estimated to be 2.5\%~\cite{CMS-PAS-LUMI-2017}.
\end{itemize}

Given that jets produced in \ttbar or \cPqt\PW\ events, regardless of the jet multiplicity of the event, are expected
to belong to the same kinematical regime, JES, \cPqb\ tagging efficiency and misidentification rate are each covered by one
single nuisance parameter.

\subsection{Modeling uncertainties}
It is important for the measurement that the modeling of the \cPqt\PW\ signal and \ttbar background events is well understood. The
impact of the theoretical assumptions in the modeling is determined
by building the templates with dedicated simulation samples of \cPqt\PW\ and \ttbar events.
These samples are produced by varying the parameters from those of the standard \POWHEG+\PYTHIA simulations.

The uncertainty in the modeling of the hard-production process is assessed by changing independently $\mu_\mathrm{R}$
and $\mu_\mathrm{F}$ in the \POWHEG sample by factors of 2 and 0.5 relative to their common nominal value,
which is set in \POWHEG to $\mu_\mathrm{R} = \mu_\mathrm{F} = \sqrt{\smash[b]{m^2_{\cPqt} + p^2_{\mathrm{T},\cPqt}}}$, where
$p_{\mathrm{T},\cPqt}$ denotes the transverse momentum of the top quark in the \ttbar rest frame.

To account for the parton shower (PS) and fragmentation uncertainties, different effects are studied:
\begin{itemize}
   \item Underlying event: \PYTHIA parameters that are tuned to the measurements of the underlying event~\cite{CMS-PAS-TOP-16-021,Skands:2014pea},
   to account for non-perturbative QCD effects, are varied up and down within their uncertainties in simulated \ttbar events.
   \item Matrix element/PS matching: the uncertainty in the combination of the matrix-element calculation with the parton shower in simulated \ttbar events
is estimated from the variation of the \POWHEG parameter $h_{\text{damp}} = 1.58^{+0.66}_{-0.59}\ m_\cPqt$~\cite{CMS-PAS-TOP-16-021}, which
regulates the damping of real emissions in the NLO calculation when matching to the PS~\cite{Skands:2014pea}.
   \item Initial- (final-) state radiation scale: the PS scale used for the simulation of the initial- (final-) state radiation
    is varied up and down by a factor of two. These variations are motivated by the uncertainties in the PS tuning~\cite{Skands:2014pea}.
   \item Color reconnection: the effect of multiple parton interactions and the parameterization of color reconnection
have been studied in Ref.~\cite{CMS-PAS-TOP-16-021} and are varied accordingly in simulated \ttbar events. In addition, we use a simulation including color reconnection of early resonant decays.  The uncertainties that arise from ambiguities in modeling color-reconnection
effects are estimated by comparing the default model in \PYTHIA with two
alternative models of color reconnection, a model with string formation beyond leading
color~\cite{Christiansen:2015yqa} and a model in which the gluons can be moved to another
string~\cite{Argyropoulos:2014zoa}. All models are tuned to measurements of the underlying
event~\cite{CMS-PAS-TOP-16-021,Skands:2014pea}.
The largest variation in each bin with respect to the nominal yield is taken as the systematic uncertainty.
   \item Semileptonic \PB\ hadron decays: the semileptonic \PB\ hadron branching fraction is varied
depending on the differences between the \PYTHIA semileptonic branching fractions for \PBz, \PBp, $\PBs^{\mathrm 0}$ and \PGLb and the Particle Data Group values~\cite{PDG2016}.
   \item \PB\ hadron fragmentation: the fragmentation into \PB\ hadrons is varied within the uncertainties of the Bowler--Lund
   fragmentation function~\cite{Bowler:1981sb} tuned to data measured by the ALEPH~\cite{Heister:2001jg} and DELPHI~\cite{DELPHI:2011aa} Collaborations. In addition, the
   difference between the Bowler--Lund and Peterson~\cite{PhysRevD.27.105} fragmentation functions is included. The largest
   variation in each bin with respect to the nominal yield is taken as the systematic uncertainty.
\end{itemize}

The uncertainty from the choice of PDFs is determined by reweighting the sample of simulated \ttbar events according to the
100 NNPDF3.0 replicas~\cite{Ball:2014uwa}. For each bin, the root-mean-square of the variation in the acceptance for all the
PDF sets is taken as an uncertainty. In order not to loose robustness in the fit, a single nuisance parameter is used.

Additionally, the difference between the DS and DR schemes is taken as a source of systematic uncertainty in the signal.

Finally, in order to extract the inclusive cross section from the measurement in the visible phase space, an extrapolation from
the visible to the total phase space is needed. This avoids constraining shape-related systematic uncertainties outside the
observable phase space (which enter the fit as normalizations). This extrapolation is made by determining the signal acceptance
from simulation. The effect of the signal modeling uncertainties in the acceptance is taken into account as an additional source
of systematic uncertainty uncorrelated with all the effects described above and added in quadrature to the total uncertainty
obtained in the fit.

Measurements of the differential cross section for top quark pair production have shown that the momentum of the top quark is softer 
than predicted by the \POWHEG simulation~\cite{Sirunyan:2017mzl,Khachatryan:2016mnb}. The effect of this mismodelling of the \pT 
spectrum was estimated by reweighting the simulation, and found to have a negligible effect.
The difference in the predictions of the NLO generators \POWHEG and \MGvATNLO for tW and \ttbar production, where both use \PYTHIA 
for hadronization, fragmentation, and additional radiation description, was estimated and found to be negligible with respect to 
the modeling uncertainties already assigned.

\subsection{Background normalization uncertainties}
For \ttbar a normalization uncertainty of 5\% is used. This takes into account effects coming from $\mu_\mathrm{R}$ and
$\mu_\mathrm{F}$ scales, PDFs and \alpS in the NNLO calculation~\cite{Czakon:2013goa}.
For DY and non-\PW/\cPZ\ backgrounds, a normalization uncertainty of $\pm$50\% is assumed. This value is motivated by the
precision of estimation methods using control regions in data, which are found to be compatible with the predictions from
the simulation. For $\ttbar$V and VV backgrounds, an uncertainty of 50\% is also used. This value reflects the uncertainties
in the corresponding predicted cross sections but is increased to account for the uncertainties due to the extrapolation of
the inclusive cross section into the phase space used in the analysis.
The overall uncertainty is not changed significantly by varying this uncertainty.

\section{Results}
\label{sec:fit}
The \cPqt\PW\ signal-strength parameter that results in the best fit to the data is \muObs,
corresponding to a measured cross section of \resultxsecmain, consistent with the SM expectations.

\begin{figure}[htbp!]
  \centering
    \includegraphics[width=0.49\textwidth]{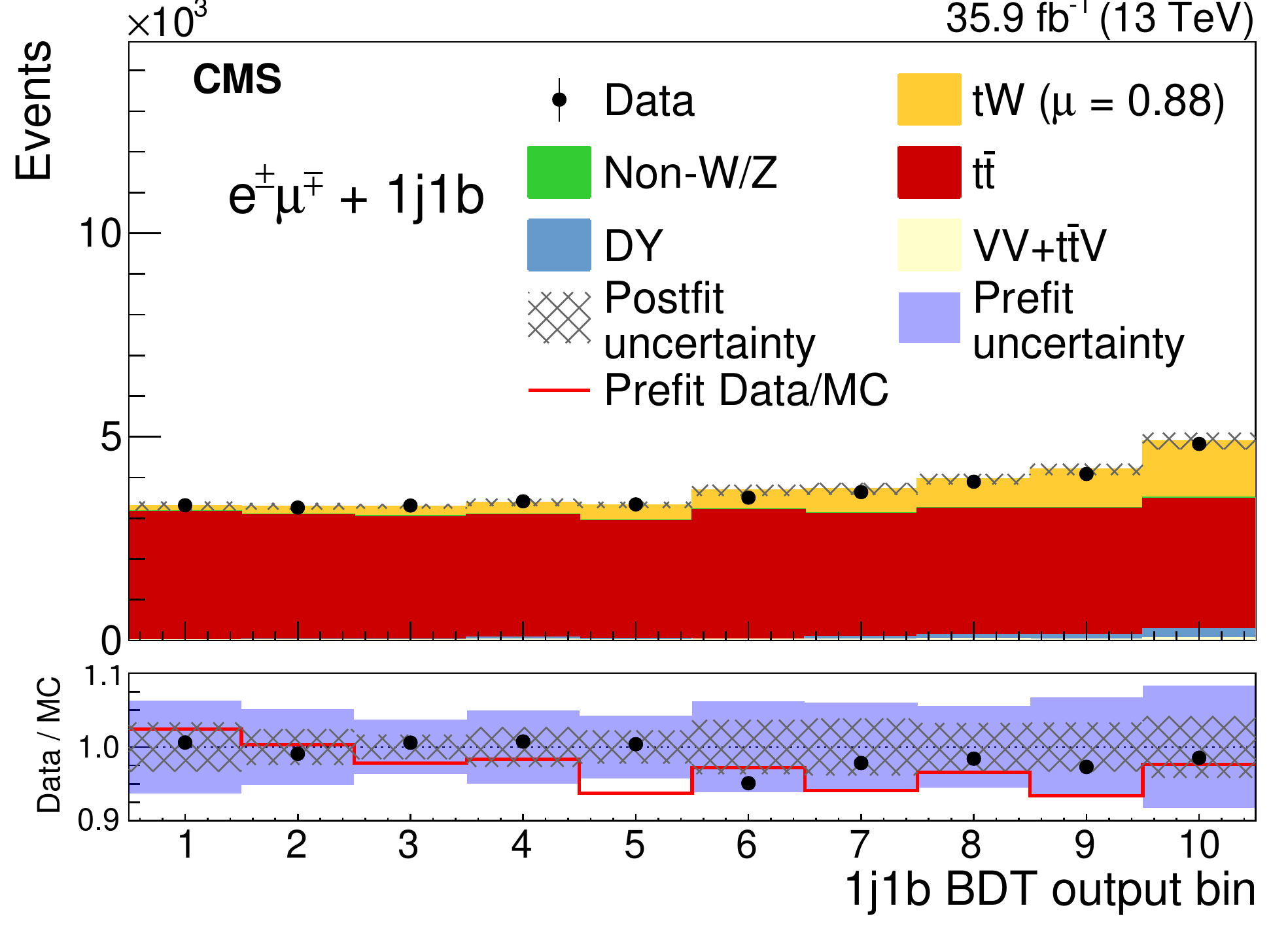}
    \includegraphics[width=0.49\textwidth]{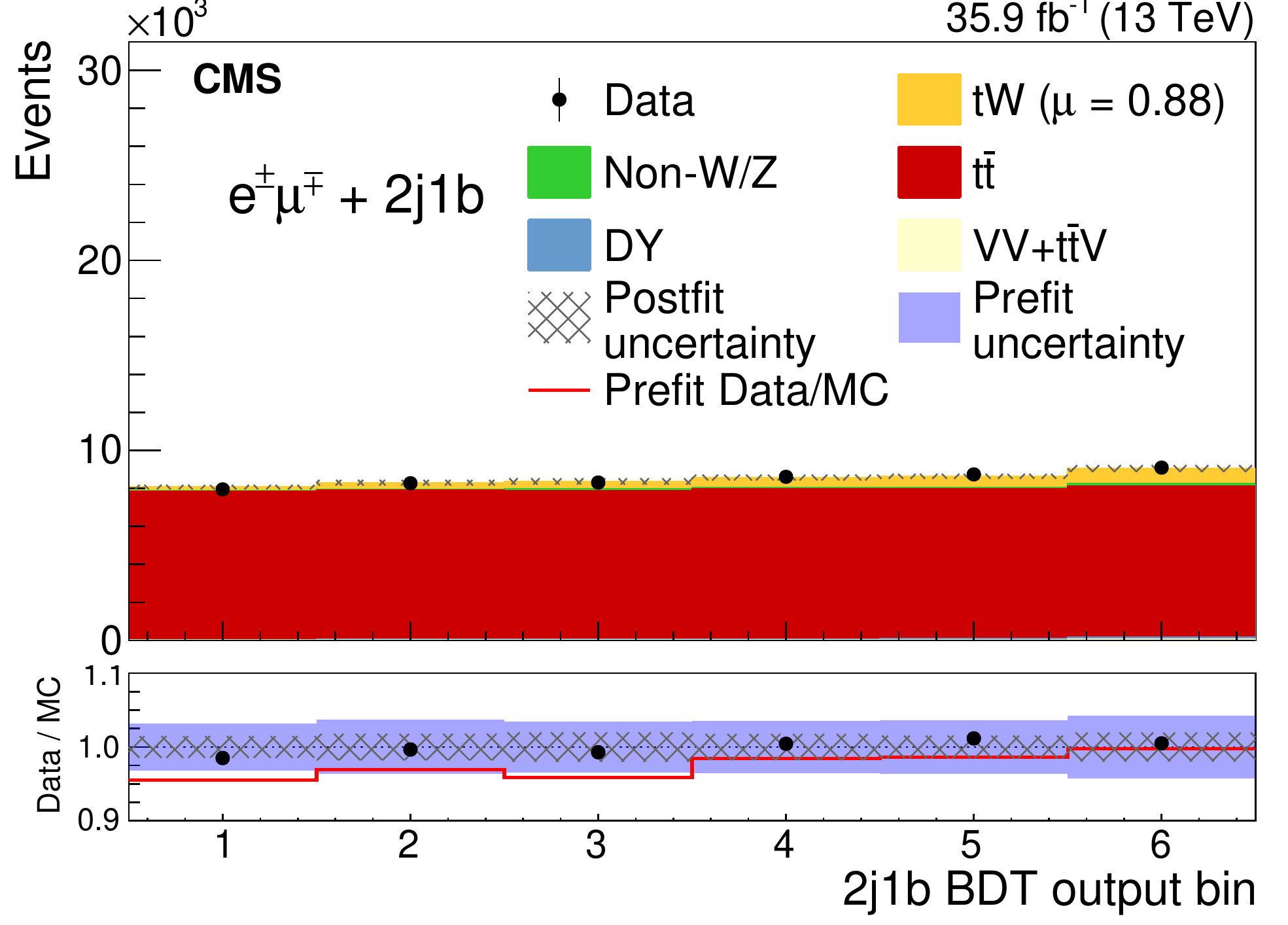} \\
    \includegraphics[width=0.49\textwidth]{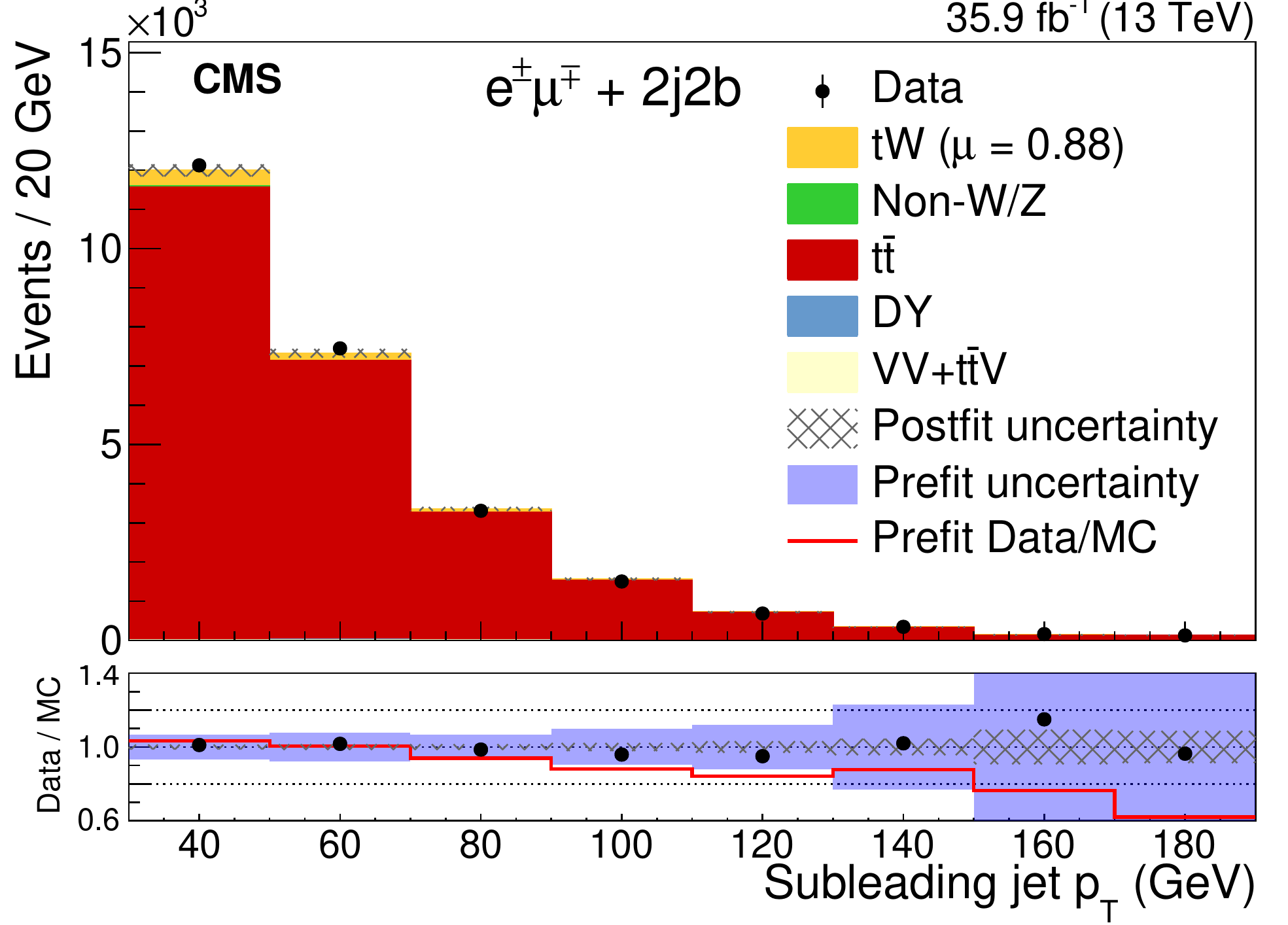}
      \caption{Comparison of the BDT output in the 1j1b (upper left) and 2j1b (upper right) regions and the \pT of the subleading jet
      in the 2j2b region (lower) distributions after the fit is performed for the observed data and simulated events. The
      error band includes the statistical and systematic uncertainties. The bottom of each panel shows the ratios of data to the prediction from simulations (line) and from the fit (dots), together with their corresponding uncertainties (solid and hatched band, respectively).}
      \label{fig:fig_outputBDTdataPostFit}
\end{figure}

Comparisons of the final distributions of the BDT discriminants in the 1j1b and 2j1b regions, as well as the
distribution of the subleading jet \pt in the 2j2b region for data and simulated events, are shown in Fig.~\ref{fig:fig_outputBDTdataPostFit}.

The number of expected events for signal and \ttbar obtained before the fit (prefit) and after the fit (postfit) are shown in Table~\ref{tab:yields_fit}.

\begin{table}[h]
\centering \topcaption{Number of expected prefit and postfit signal and \ttbar background events.}
\label{tab:yields_fit}
\begin{tabular}{lcccc}
          &\multicolumn{2}{c}{Prefit}                & \multicolumn{2}{c}{Postfit} \\
\hline
Region    & \cPqt\PW\             &  \ttbar          &  \cPqt\PW\            &   \ttbar          \\
\hline
1j1b      & 6147 $\pm$ 442 & 30622 $\pm$ 1862 & 5440 $\pm$ 604 &   30592 $\pm$ 582 \\
2j1b      & 3125 $\pm$ 294 & 48484 $\pm$ 1984 & 2888 $\pm$ 321 &   47436 $\pm$ 612 \\
2j2b      &  725 $\pm$  85 & 25052 $\pm$ 2411 &  719 $\pm$  88 &   25114 $\pm$ 281 \\
\end{tabular}
\end{table}

Several nuisance parameters (JES, \ttbar modeling) are significantly constrained due to their effect on
the jet multiplicity and the input distributions used in the fit. The \ttbar normalization is also constrained due to the large
presence of \ttbar in the different regions.

The impact of each source of systematic uncertainty in the fit, shown in Table~\ref{tab:systs_fit}, is evaluated by performing
the fit, fixing the rest of the nuisance parameters to their postfit value. We take the difference in quadrature between the
uncertainty of the fit with all the nuisance parameters except the one under study fixed to the postfit value, and the uncertainty
of the fit with all the nuisances fixed to the postfit value. The uncertainties in the luminosity and in the trigger and lepton
efficiencies lead to uncertainties in the background, which is dominant in all bins of the fit. Therefore, these uncertainties
make a sizable contribution to the uncertainty in the final measurement.

\begin{table}[htbp]
\centering \topcaption{Estimation of the effect on the signal strength of each source of uncertainty in the fit. Experimental and modeling uncertainties affect both the rate and the shape of the templates while background normalization uncertainties affect only the rate.}
\label{tab:systs_fit}
\begin{tabular}{lc}
          Source & Uncertainty (\%)  \\ \hline
Experimental &   \\
\hspace*{0.5cm} Trigger efficiencies  & 2.7 \\
\hspace*{0.5cm}         Electron efficiencies  & 3.2 \\
\hspace*{0.5cm}             Muon efficiencies  & 3.1 \\
\hspace*{0.5cm}                           JES  & 3.2 \\
\hspace*{0.5cm}         Jet energy resolution  & 1.8 \\
\hspace*{0.5cm}     \cPqb\ tagging efficiency  & 1.4 \\
\hspace*{0.5cm}                   Mistag rate  & 0.2 \\
\hspace*{0.5cm}                        Pileup  & 3.3 \\[\cmsTabSkip]
Modeling & \\
\hspace*{0.5cm}  \ttbar\ $\mu_\mathrm{R}$ and $\mu_\mathrm{F}$ scales & 2.5 \\
\hspace*{0.5cm}  \cPqt\PW\ $\mu_\mathrm{R}$ and $\mu_\mathrm{F}$ scales & 0.9 \\
\hspace*{0.5cm}             Underlying event  & 0.4 \\
\hspace*{0.5cm}               Matrix element/PS matching  & 1.8 \\
\hspace*{0.5cm}      Initial-state radiation  & 0.8 \\
\hspace*{0.5cm}        Final-state radiation  & 0.8 \\
\hspace*{0.5cm}            Color reconnection & 2.0 \\
\hspace*{0.5cm}           \PB\ fragmentation  & 1.9 \\
\hspace*{0.5cm}      Semileptonic \PB\ decay  & 1.5 \\
\hspace*{0.5cm}                         PDFs  & 1.5 \\
\hspace*{0.5cm}                        DR-DS  & 1.3 \\[\cmsTabSkip]
Background normalization & \\
\hspace*{0.5cm}                      \ttbar   & 2.8 \\
\hspace*{0.5cm}                          VV   & 0.4 \\
\hspace*{0.5cm}                  Drell--Yan   & 1.1 \\
\hspace*{0.5cm}         Non-\PW/\cPZ\ leptons & 1.6 \\
\hspace*{0.5cm}                    $\ttbar$V  & 0.1 \\[\cmsTabSkip]
                        MC finite sample size & 1.6 \\[\cmsTabSkip]
               Full phase space extrapolation & 2.9 \\[\cmsTabSkip]
                             Total systematic & \multirow{2}{*}{10.1}  \\
            (excluding integrated luminosity) &     \\[\cmsTabSkip]
                        Integrated luminosity & 3.3 \\[\cmsTabSkip]
                                  Statistical & 2.8 \\[\cmsTabSkip]
                                        Total & 11.1 \\
\end{tabular}
\end{table}

\section{Summary}
\label{sec:summary}
The data recorded by CMS at 13\TeV, corresponding to an integrated luminosity of
\lumivwunc, are used to measure the \cPqt\PW\ production cross section in the \emu channel, classifying the events in terms of the number
of jets and jets originating from bottom quarks.
The signal is measured using a maximum likelihood fit to the distribution of boosted decision tree discriminants in two of the categories,
and to the \pt distribution of the second jet with highest \pt in a third category.
The measured cross section for \cPqt\PW\ production is found to be \resultxsecmain, achieving a relative uncertainty of 11\%.
This is the first measurement of this process by the CMS Collaboration at $\roots= 13\TeV$. The measured cross section
is in agreement with the standard model prediction of \tWxsectheo and with a similar measurement by the ATLAS Collaboration~\cite{Aaboud:2016lpj}.

\begin{acknowledgments}
\hyphenation{Bundes-ministerium Forschungs-gemeinschaft Forschungs-zentren Rachada-pisek} We congratulate our colleagues in the CERN accelerator departments for the excellent performance of the LHC and thank the technical and administrative staffs at CERN and at other CMS institutes for their contributions to the success of the CMS effort. In addition, we gratefully acknowledge the computing centres and personnel of the Worldwide LHC Computing Grid for delivering so effectively the computing infrastructure essential to our analyses. Finally, we acknowledge the enduring support for the construction and operation of the LHC and the CMS detector provided by the following funding agencies: the Austrian Federal Ministry of Science, Research and Economy and the Austrian Science Fund; the Belgian Fonds de la Recherche Scientifique, and Fonds voor Wetenschappelijk Onderzoek; the Brazilian Funding Agencies (CNPq, CAPES, FAPERJ, and FAPESP); the Bulgarian Ministry of Education and Science; CERN; the Chinese Academy of Sciences, Ministry of Science and Technology, and National Natural Science Foundation of China; the Colombian Funding Agency (COLCIENCIAS); the Croatian Ministry of Science, Education and Sport, and the Croatian Science Foundation; the Research Promotion Foundation, Cyprus; the Secretariat for Higher Education, Science, Technology and Innovation, Ecuador; the Ministry of Education and Research, Estonian Research Council via IUT23-4 and IUT23-6 and European Regional Development Fund, Estonia; the Academy of Finland, Finnish Ministry of Education and Culture, and Helsinki Institute of Physics; the Institut National de Physique Nucl\'eaire et de Physique des Particules~/~CNRS, and Commissariat \`a l'\'Energie Atomique et aux \'Energies Alternatives~/~CEA, France; the Bundesministerium f\"ur Bildung und Forschung, Deutsche Forschungsgemeinschaft, and Helmholtz-Gemeinschaft Deutscher Forschungszentren, Germany; the General Secretariat for Research and Technology, Greece; the National Scientific Research Foundation, and National Innovation Office, Hungary; the Department of Atomic Energy and the Department of Science and Technology, India; the Institute for Studies in Theoretical Physics and Mathematics, Iran; the Science Foundation, Ireland; the Istituto Nazionale di Fisica Nucleare, Italy; the Ministry of Science, ICT and Future Planning, and National Research Foundation (NRF), Republic of Korea; the Lithuanian Academy of Sciences; the Ministry of Education, and University of Malaya (Malaysia); the Mexican Funding Agencies (BUAP, CINVESTAV, CONACYT, LNS, SEP, and UASLP-FAI); the Ministry of Business, Innovation and Employment, New Zealand; the Pakistan Atomic Energy Commission; the Ministry of Science and Higher Education and the National Science Centre, Poland; the Funda\c{c}\~ao para a Ci\^encia e a Tecnologia, Portugal; JINR, Dubna; the Ministry of Education and Science of the Russian Federation, the Federal Agency of Atomic Energy of the Russian Federation, Russian Academy of Sciences, the Russian Foundation for Basic Research and the Russian Competitiveness Program of NRNU ``MEPhI"; the Ministry of Education, Science and Technological Development of Serbia; the Secretar\'{\i}a de Estado de Investigaci\'on, Desarrollo e Innovaci\'on, Programa Consolider-Ingenio 2010, Plan de Ciencia, Tecnolog\'{i}a e Innovaci\'on 2013-2017 del Principado de Asturias and Fondo Europeo de Desarrollo Regional, Spain; the Swiss Funding Agencies (ETH Board, ETH Zurich, PSI, SNF, UniZH, Canton Zurich, and SER); the Ministry of Science and Technology, Taipei; the Thailand Center of Excellence in Physics, the Institute for the Promotion of Teaching Science and Technology of Thailand, Special Task Force for Activating Research and the National Science and Technology Development Agency of Thailand; the Scientific and Technical Research Council of Turkey, and Turkish Atomic Energy Authority; the National Academy of Sciences of Ukraine, and State Fund for Fundamental Researches, Ukraine; the Science and Technology Facilities Council, UK; the US Department of Energy, and the US National Science Foundation.

Individuals have received support from the Marie-Curie programme and the European Research Council and Horizon 2020 Grant, contract No. 675440 (European Union); the Leventis Foundation; the A. P. Sloan Foundation; the Alexander von Humboldt Foundation; the Belgian Federal Science Policy Office; the Fonds pour la Formation \`a la Recherche dans l'Industrie et dans l'Agriculture (FRIA-Belgium); the Agentschap voor Innovatie door Wetenschap en Technologie (IWT-Belgium); the Ministry of Education, Youth and Sports (MEYS) of the Czech Republic; the Council of Scientific and Industrial Research, India; the HOMING PLUS programme of the Foundation for Polish Science, cofinanced from European Union, Regional Development Fund, the Mobility Plus programme of the Ministry of Science and Higher Education, the National Science Center (Poland), contracts Harmonia 2014/14/M/ST2/00428, Opus 2014/13/B/ST2/02543, 2014/15/B/ST2/03998, and 2015/19/B/ST2/02861, Sonata-bis 2012/07/E/ST2/01406; the National Priorities Research Program by Qatar National Research Fund; the Programa Severo Ochoa del Principado de Asturias; the Thalis and Aristeia programmes cofinanced by EU-ESF and the Greek NSRF; the Rachadapisek Sompot Fund for Postdoctoral Fellowship, Chulalongkorn University and the Chulalongkorn Academic into Its 2nd Century Project Advancement Project (Thailand); the Welch Foundation, contract C-1845; and the Weston Havens Foundation (USA).
\end{acknowledgments}

\bibliography{auto_generated}

\cleardoublepage \appendix\section{The CMS Collaboration \label{app:collab}}\begin{sloppypar}\hyphenpenalty=5000\widowpenalty=500\clubpenalty=5000\input{TOP-17-018-authorlist.tex}\end{sloppypar}
\end{document}

%% file: TOP-17-018-authorlist.tex
\vskip\cmsinstskip
\textbf{Yerevan Physics Institute, Yerevan, Armenia}\\*[0pt]
A.M.~Sirunyan, A.~Tumasyan
\vskip\cmsinstskip
\textbf{Institut f\"{u}r Hochenergiephysik, Wien, Austria}\\*[0pt]
W.~Adam, F.~Ambrogi, E.~Asilar, T.~Bergauer, J.~Brandstetter, E.~Brondolin, M.~Dragicevic, J.~Er\"{o}, A.~Escalante~Del~Valle, M.~Flechl, M.~Friedl, V.M.~Ghete, J.~Hrubec, M.~Jeitler\cmsAuthorMark{1}, N.~Krammer, I.~Kr\"{a}tschmer, D.~Liko, T.~Madlener, I.~Mikulec, N.~Rad, H.~Rohringer, J.~Schieck\cmsAuthorMark{1}, R.~Sch\"{o}fbeck, M.~Spanring, D.~Spitzbart, A.~Taurok, W.~Waltenberger, J.~Wittmann, C.-E.~Wulz\cmsAuthorMark{1}, M.~Zarucki
% ...
\vskip\cmsinstskip
% ...
1:  Also at Vienna University of Technology, Vienna, Austria\\
% ...
